\documentclass[aps,prl,twocolumn,floatfix,footinbib,showpacs,superscriptaddress]{revtex4-1}
\usepackage{graphicx}
\usepackage{amsfonts,amsmath,amssymb}
\usepackage{amsthm}
\usepackage{dsfont,bm}
\usepackage{color}
\usepackage{soul} 
\usepackage{amsbsy}
\usepackage[colorlinks=true,linkcolor=blue,pagecolor=blue,filecolor=blue,menucolor=blue,urlcolor=blue,citecolor=blue,anchorcolor=blue]{hyperref}%
\usepackage{multirow}
\usepackage{mathrsfs}
\usepackage{txfonts}
\usepackage{amsbsy}
\usepackage{times} 
\usepackage{dcolumn}

\newcommand{\ua}{\uparrow}
\newcommand{\da}{\downarrow}

\usepackage{lineno}

\begin{document}

\title{Josephson effect in graphene bilayers with adjustable relative displacement}

\author{Mohammad Alidoust}
\affiliation{Department of Physics, NTNU Norwegian University of Science and Technology, N-7491 Trondheim, Norway}
\author{Antti-Pekka Jauho}
\affiliation{Center for Nanostructured Graphene (CNG), DTU Physics, Technical University of Denmark, DK-2800 Kongens Lyngby, Denmark}
\author{Jaakko Akola} 
\affiliation{Department of Physics, Norwegian University of Science and Technology, N-7491 Trondheim, Norway}
\affiliation{Computational Physics Laboratory, Tampere University, P.O. Box 692, FI-33014 Tampere, Finland}

\begin{abstract}
The Josephson current is investigated in a superconducting graphene bilayer where pristine graphene sheets can make in-plane or out-of-plane displacements with respect to each other. The superconductivity can be of an intrinsic nature, or due to a proximity effect. The results demonstrate that the supercurrent responds qualitatively differently to relative displacement if the superconductivity is due to either intralayer or interlayer spin-singlet electron-electron pairing, thus providing a tool to distinguish between the two mechanisms. Specifically, both the AA and AB stacking orders are studied with antiferromagnetic spin alignment. For the AA stacking order with intralayer and on-site pairing no current reversal is found. In contrast, the supercurrent may switch its direction as a function of the in-plane displacement and out-of-plane interlayer coupling for the cases of AA ordering with interlayer pairing and AB ordering with either intralayer or interlayer pairing. In addition to sign reversal, the Josephson signal displays many characteristic fingerprints which derive directly from the pairing mechanism. Thus, measurements of the Josephson current as a function of the graphene bilayer displacement open up the means achieve deeper insights into the superconducting pairing mechanism.    
\end{abstract}

\date{\today}

\maketitle

Two-dimensional materials have attracted enormous attention during the last decade. One of the main reasons is the unique opportunity to make these materials thinner down to atomic mono- or bilayers  \cite{RMP-2009-Neto,RMP-2008-Beenakker}, and the possibility of a subsequent stacking of the constituents at an essentially arbitrary relative displacement or twist angle that results in qualitative consequences on the microscopic and macroscopic properties of the system. A prominent example is the "magic angle" twisted bilayer graphene (tBLG) \cite{tbg1,tbg-1,tbg2,tbg3,tbg4,tbg5,tbg6,tbg9,tbg10,tbg11,tbg12,tbg14,2ph-el_exp,A.L.Sharpe,Y.Choi,Y.Jiang,Y.Xie,A.Kerelsky,A.O.Sboychakov1,A.O.Sboychakov2,A.O.Sboychakov3,N.Bultinck,tbg18,tbg19} that develops intrinsic superconductivity at $T_c\simeq 1.7$ K \cite{herrero,X.Lu,M.Yankowitz,G.Chen}. Unlike the traditional BCS scenario of superconductivity that requires a high-density of free electrons, a tBLG at a magic angle has an extremely low density of electrons ($\rm n_e\approx 10^{11} cm^{-2}$). The ratio between the electron density and superconducting critical temperature places tBLG in the high-$T_c$ part of the phase diagram of superconductors \cite{herrero,herrero2}. Recent samples show an enhanced $T_c \rm \sim 3 K$, and magnetic states in the vicinity of interaction-induced insulating states \cite{X.Lu}. These striking phenomena have raised extensive theoretical discussions on the origin and type of superconductivity in these systems \cite{teor_tbg1,teor_tbg2,teor_tbg3,teor_tbg4,teor_tbg5,teor_tbg7,teor_tbg9,teor_tbg11,teor_tbg12,teor_tbg13,teor_tbg14,teor_tbg15,teor_tbg16,teor_tbg17,teor_tbg18,teor_tbg19,teor_tbg20,teor_tbg21,teor_tbg22,teor_tbg23,Hesselmann,alidoust2019prb,F.Wu1,F.Wu2,E.F.Talantsev1,E.F.Talantsev2,M.S.Scheurer}. Nevertheless, no universally shared conclusion has been reached so far and a clear-cut picture remains elusive. 

The emergence of superconductivity adjacent to insulating states with broken spin-valley degeneracy may hint at an exotic superconducting pairing mechanism. One main scenario  includes various types of phonon-mediated $d$-wave pairings; $d_{x^2-y^2}, d_xd_y$, while another candidate deals with a direct spin-spin interaction \cite{teor_tbg1,teor_tbg2,teor_tbg3,teor_tbg4,teor_tbg5,teor_tbg7,teor_tbg9,teor_tbg11,teor_tbg12,teor_tbg13,teor_tbg14,teor_tbg15,teor_tbg16,teor_tbg17,teor_tbg18,teor_tbg19,teor_tbg20,teor_tbg21,teor_tbg22,teor_tbg23,Hesselmann,alidoust2019prb,F.Wu1,F.Wu2,E.F.Talantsev1,E.F.Talantsev2,M.S.Scheurer}. The experimental phenomenology in tBLG is similar to the high-$ T_c$  superconductivity observed in cuprates, and thus the putative displacement- or twist-induced $d$-wave symmetry in tBLG might offer some useful hints towards the cuprate problem. As Raman spectroscopy experiments have revealed, BLG systems can support two phonon modes \cite{tbg2,tbg3,tbg4,tbg5,tbg6,tbg9,tbg10,tbg11,tbg12,tbg14,2ph-el_exp}, one corresponding to intralayer vibrations and the other to interlayer vibrations. In this spirit, a two-gap superconductivity has been proposed and studied for displaced and commensurate twisted BLG systems \cite{alidoust2019prb,E.F.Talantsev1,E.F.Talantsev2}. It was found that, depending on the original stacking order of BLG (being either AA or AB) and the magnitude of the chemical potential, a small in-plane displacement of the pristine layers with respect to each other can drive $s$-wave and $p$-wave pairing symmetries into $d$-wave and $f$-wave symmetry classes, respectively, and increase the superconducting critical temperature \cite{alidoust2019prb}.

\begin{figure}[b!]
\includegraphics[clip, trim=0.0cm 0.0cm 0.0cm 0.1cm, width=8.50cm,height=6.5cm]{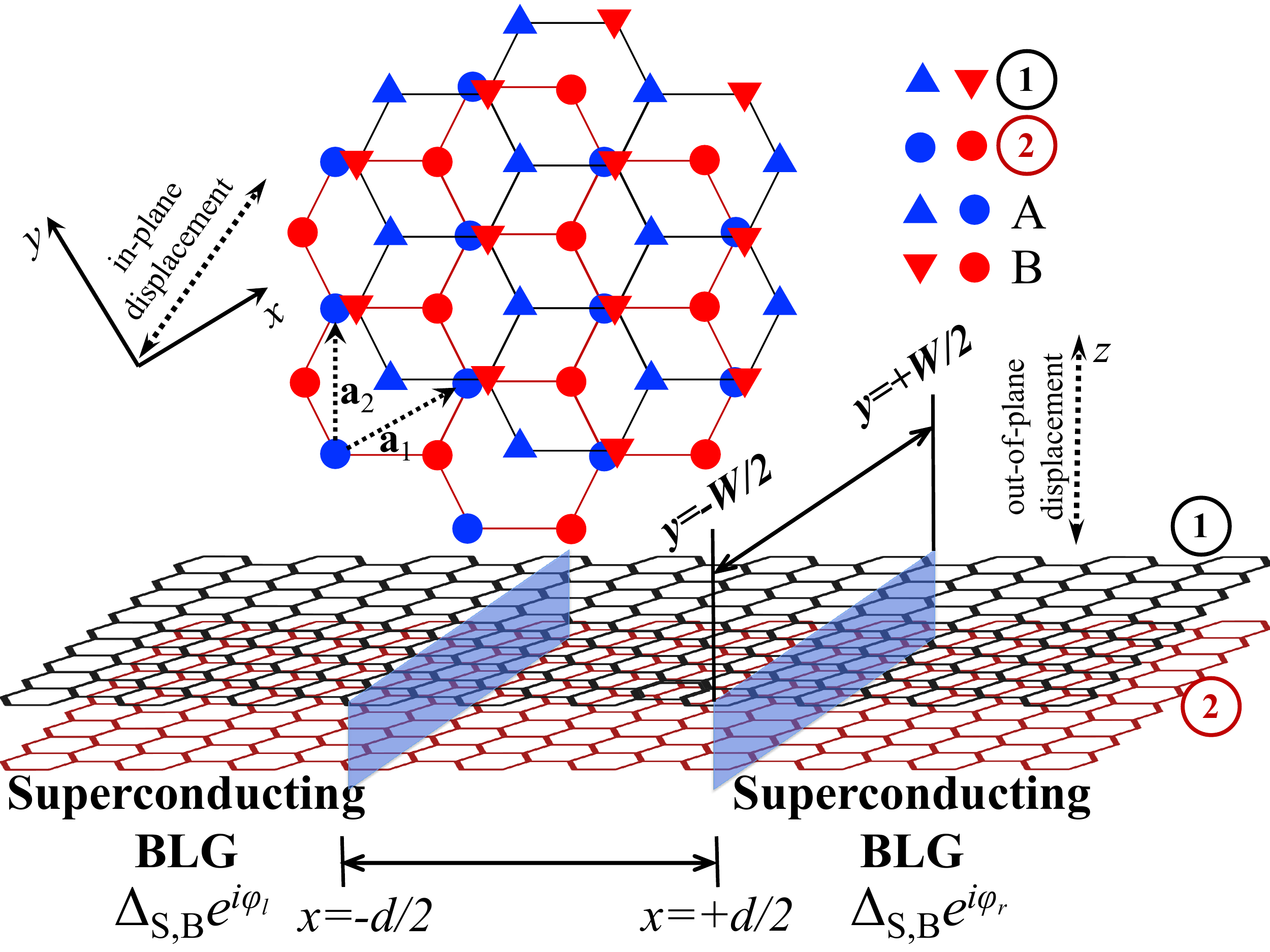}
\caption{\label{fig1} (Color online). The SNS BLG configuration with AB stacking order. The top ($1$) and bottom ($2$) pristine graphene layers are coupled and can make in-plane and out-of-plane movements with respect to each other. The atomic sites of the top and bottom layers are shown by circles and triangles, while sublattices A and B in both layers are blue and red, respectively.  The primitive lattice vectors are $\rm {\bf a}_1=a(1,0)$ and $\rm {\bf a}_2=a(1,\sqrt{3})/2$ with $\rm a=2.46\AA$. The sublattice positions in real space are $\rm \boldsymbol{\nu}_{1,A}= (0, 0)$ and $\rm \boldsymbol{\nu}_{1,B}= 2({\bf a}_1 + {\bf a}_2)/3$. In AA stacking, $\rm \boldsymbol{\nu}_{2,A}=(0, 0)$ and $\rm \boldsymbol{\nu}_{2,B}=2({\bf a}_1 + {\bf a}_2)/3$, in AB stacking $\rm \boldsymbol{\nu}_{2,A}=({\bf a}_1 + {\bf a}_2)/3$ and $\rm \boldsymbol{\nu}_{2,B}=(0, 0)$, and AC stacking $\rm \boldsymbol{\nu}_{2,A}=2({\bf a}_1 + {\bf a}_2)/3$ and $\rm \boldsymbol{\nu}_{2,B}=({\bf a}_1 + {\bf a}_2)/3$. The light blue planes show the boundaries between the superconducting and normal segments. The two superconducting parts have different macroscopic phases $\varphi_{l,r}$, and the superconductivity can be of either intralayer $(\Delta_\text{S})$ or interlayer $(\Delta_\text{B})$ origin. The two-dimensional system with width $W$ resides in the $xy$-plane and the two superconducting parts are separated by a distance $d$ in the $x$ direction. The lengths are in units of the the superconducting coherence length $\xi_\text{S}=\hbar v_{F}/\Delta$ ($v_{F}$ is the Fermi velocity) and the energies in units of the superconducting gap $\Delta$, respectively.}
\end{figure}

\begin{figure*}[t]
\includegraphics[ width=17.50cm,height=7.3cm]{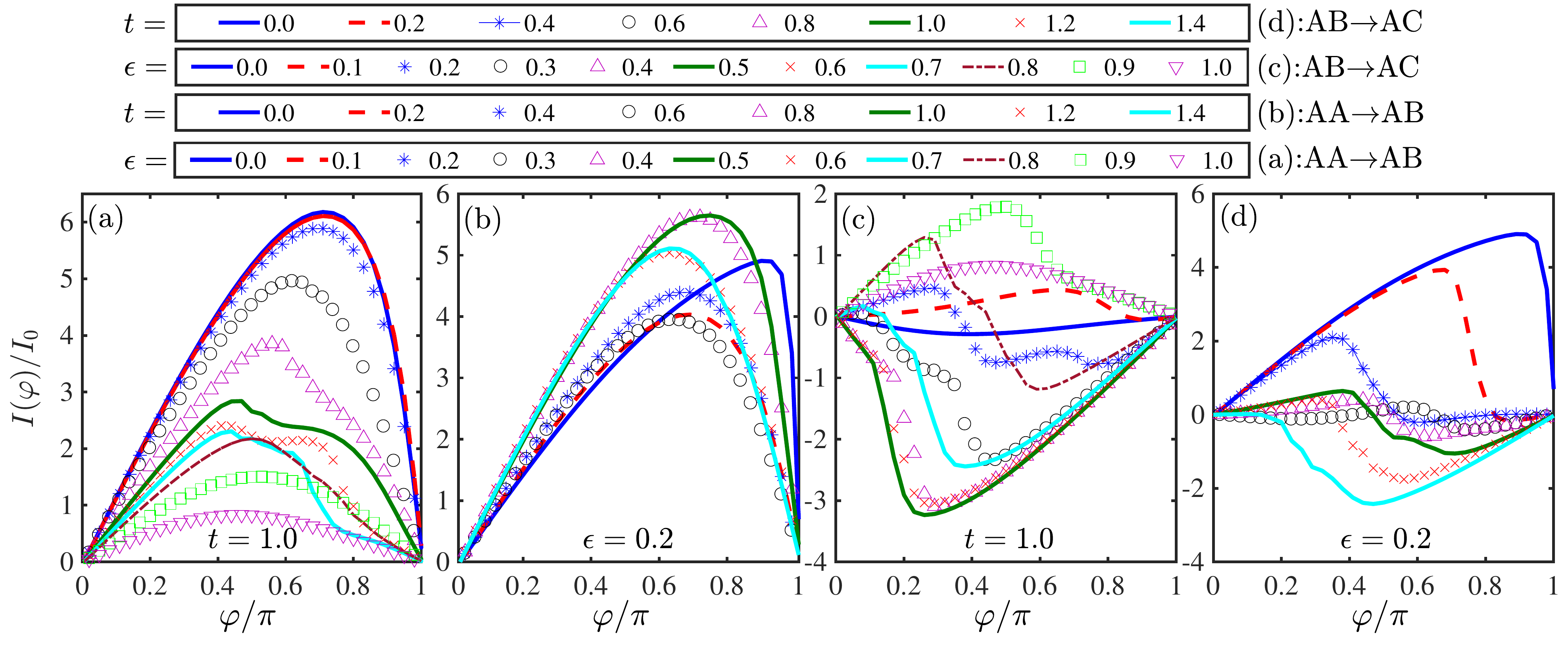}
\caption{\label{fig2} (Color online). The normalized Josephson current by $I_0=2e|\Delta|W$ (in which $e>0$ is the electron charge) for an opposite-spin intralayer electron-electron coupling. In (a) and (b), the initial stacking order is AA while in (c) and (d) the initial ordering is AB. In (a) and (c), the stacking order transforms into AB and AC, respectively, by varying $\epsilon$. In (b) and (d), a fixed displacement $\epsilon=0.2$ is considered and the coupling strength $t$ is varied. The chemical potential is set $\mu=1.5\Delta$ (this is a representative value).}
\end{figure*}

It is apparent that any experiment that yields additional information about the pairing mechanism in a superconducting BLG is highly desirable. In this Rapid Communication, we adopt an effective Hamiltonian model where the interaction between the two pristine graphene layers is described by a coupling matrix \cite{ray1,ray2,ray_slip}. When one of the pristine graphene layers is subjected to an in-plane displacement or commensurate twist, all corresponding modifications are encoded into the coupling matrix. We consider a Josephson junction made of a BLG, shown in Fig. \ref{fig1}, and study the influence of the layers' relative displacement on supercurrent flow in the presence of intralayer or interlayer superconducting pairings. In contrast to on-site spin-singlet superconductivity or intralayer unequal-spin pairing with an AA initial stacking order, in an AB stacking the displacement causes supercurrent reversal and induces higher order harmonics at the crossover points, both for unequal-spin intralayer or interlayer pairing. The latter signatures appear in AA ordering with interlayer superconductivity as well. As the supercurrent is one of the most directly accessible quantities in experiments \cite{X.Lu,M.Yankowitz}, our findings offer a novel probe for characterizing the mechanism underlying superconductivity in a BLG system. Furthermore, according to our predictions, the supercurrent reversals can potentially be utilized in future BLG superconducting memory devices. We note that the unit cell of BLG close to a magic angle becomes very large so that a precise effective Hamiltonian model at this limit seems impractical. Nevertheless, the antiferromagnetic electron-electron pairing scenario we incorporate in our model is similar to earlier theory studies that demonstrated this kind of pairing as the most energetically stable scenario in magic-angle BLG \cite{T.Huang}. It is also consistent with recent experiments that showed a nontrivial hysteresis diagram in magic-angle BLG, suggestive of a ferromagnetic phase of BLG subject to an external magnetic field \cite{A.L.Sharpe}.       
\par
The low-energy effective Hamiltonian that describes a BLG with an in-plane displacement between the pristine layers reads 
\begin{eqnarray}\label{Hamil1}
H= &&\int \frac{d\textbf{k}}{(2\pi)^2}\hat{\psi}^\dag(\textbf{k})H(\textbf{k})\hat{\psi}(\textbf{k})
=\int \frac{d\textbf{k}}{(2\pi)^2}\hat{\psi}^\dag(\textbf{k})\\ &&\times \Big\{  H_1({\bf k})\rho_1+H_2({\bf k})\rho_2 + \tilde{\bm T}(\textbf{k})\rho_+ + \tilde{\bm T}^\dag(\textbf{k})\rho_-   \Big\}\hat{\psi}(\textbf{k}).\nonumber\label{heff}
\end{eqnarray}
The layer Pauli matrices are denoted by $\rho_{0,z,x,y}$ and $2\rho_1=\rho_0+\rho_z$, $2\rho_2=\rho_0-\rho_z$, $2\rho_+=\rho_x+i\rho_y$, $2\rho_-=\rho_x-i\rho_y$. The top ($1$) and bottom ($2$) pristine graphene in Fig. \ref{fig1} are coupled by $\tilde{\bm T}(\textbf{k})$ \cite{ray1,ray2,ray_slip}. The particles in the pristine sheets are governed by $H_{1,2}({\bf k})=\hbar v_\textbf{F}{\bf k}\cdot {\bm \sigma}$. Here ${\bm \sigma}=(\sigma_x,\sigma_y)$ are pseudo-spin Pauli matrices. The associated field operator can be expressed by $\hat{\psi}^\dag(\textbf{k})=(\psi_{1\ua}^\dag,\psi_{1\da}^\dag,\psi_{2\ua}^\dag,\psi_{2\da}^\dag)$ where $\uparrow$ ($\downarrow$) stands for the pseudo-spin up -sublattice A- (pseudo-spin down -sublattice B-) \cite{RMP-2009-Neto,RMP-2008-Beenakker}. A detailed description of the notation used here can be found in Ref. \cite{alidoust2019prb}. The lattice vectors ${\bm b}_1 = 2a^{-1}\pi(1,-1/\sqrt{3}), {\bm b}_2 = 2a^{-1}\pi(0,2/\sqrt{3})$ span the reciprocal space of BLG and in the low-energy regime the pertinent Hamiltonian can be obtained by an expansion of the tight-binding Hamiltonian around the points, $\textbf{K}_j=\textbf{K}_0+\textbf{G}_j$, with $\textbf{K}_0=(2{\bm b}_1+{\bm b}_2)/3, \textbf{G}_0={\bm 0}, \textbf{G}_1=-{\bm b}_1, \textbf{G}_2=-{\bm b}_1-{\bm b}_2$. The coupling matrix can be expressed as
\begin{equation}\label{tmatrix}
\tilde{\bm T}(\textbf{k})=\sum_{j=0,1,2}\text{M}_j^\text{XX} \frac{t_{\perp}(\textbf{K}_j+\textbf{k})}{3}e^{i(\textbf{K}_j+\textbf{k})(\textbf{u}_{2}-\textbf{u}_{1})}.
\end{equation}
Here, the interlayer hopping amplitude is given by $t_{\perp}({\bf q})=V_u^{-1}\int d{\bf r} t_{\perp}({\bf r})e^{i{\bf q}\cdot{\bf r}}$, $t_{\perp}({\bf r})=\sum_{i,j}t_{2,j}^{1,i}\Big\langle {\bf r},2\Big|c^\dag_{2,j}c_{1,i}\Big|{\bf r}',1\Big\rangle$ where $c^\dag,c$ are the quasiparticles' creation and annihilation operators, respectively. The couplings between pristine layers $1$ and $2$ are located at ${\bf r}$ and ${\bf r}'={\bf r}+{\bm \delta}$, where ${\bm \delta}$ is the distance between two hopping sites. The volume of the unit cell is $V_u$, the indices $i,j$ run over the lattice sites, $\textbf{u}_{1,2}$ vectors are the displacements of the top and bottom layers, and $\text{M}_j^\text{XX}$ encodes AA, AB, and AC orderings of the bilayer \cite{sm}.

Note that an AA stacking order can evolve into an AB structure by shifting the layer $1$ through $\textbf{u}_{1}= \epsilon({\bm a}_1 + {\bm a}_2)/3$ in Eq. (\ref{tmatrix}), as shown in Fig. \ref{fig1}. This transformation evolves the AA, AB, and AC initial orderings into AB, AC, and AA orderings, respectively, when changing $\epsilon$ from 0 to unity (see Refs. \cite{ray1,ray2,ray_slip,alidoust2019prb}). The resulting modifications of the respective band structures are presented in the Supplemental Material~\onlinecite{sm}.

In order to simulate the intralayer and interlayer superconductivity, originating from the two phonon modes (vibrations within a pristine layer and between the layers), we consider the following electron-electron coupling amplitudes; (i) $   \Delta_\text{S}\Big\langle\psi^\dag_{1\ua}\psi^\dag_{1\da}\Big\rangle+\text{H.c.}, \Delta_\text{S}\Big\langle\psi^\dag_{2\ua}\psi^\dag_{2\da}\Big\rangle+\text{H.c.}$ and (ii) $\Delta_\text{B}\Big\langle\psi^\dag_{1\ua}\psi^\dag_{2\da}\Big\rangle+\text{H.c.}$.
The amplitudes of the opposite-spin electron-electron couplings within each layer and between the layers are $\Delta_\text{S}$ and $\Delta_\text{B}$, respectively. Note that the superconducting phase in a magic-angle BLG emerges adjacent to the interaction-driven insulating states that break spin-valley degeneracy\cite{A.L.Sharpe} and suggest an unconventional pairing mechanism. Therefore, the anti-ferromagnetic model we consider here, even though it is not derived for the magic-angle of BLG, may have relevance to it \cite{herrero,X.Lu,M.Yankowitz,G.Chen}. 

In the presence of superconductivity, the low-energy Hamiltonian including particle and hole excitations reads
\begin{equation}\label{Hamil}
{\cal H} (\textbf{k})= \left( \begin{array}{cc}
H(\textbf{k})-\mu & \hat{\Delta}\\
 \hat{\Delta}^\dag & -{\cal T}H(\textbf{k}){\cal T}^{-1}+\mu
\end{array}\right),
\end{equation}
where $\hat{\Delta}$ is the $4\times 4$ superconducting gap matrix. The chemical potential is denoted by $\mu$ and the hole-block is obtained by acting with the time reversal operator $\cal T$ on the single-particle Hamiltonian $H(\textbf{k})$. The supercurrent flow across the Josephson junction shown in Fig. \ref{fig1} is evaluated directly from the definition of the current, ${\bm J} =\int \hspace{-.1cm} d{\bm r}\left\{\hat{\psi}^\dagger({\bm r}) \overrightarrow{{\cal H}}({\bm r})\hat{\psi}({\bm r})-
\hat{\psi}^\dagger({\bm r}) \overleftarrow{{\cal H}}({\bm r})\hat{\psi}({\bm r}) \right\}$.
The Hamiltonian in real space ${\cal H}\big({\bm r}\equiv (x,y,0)\big)$ can be obtained by substituting $i\textbf{ k}\equiv (\partial_x,\partial_y)$ in Eq.~(\ref{Hamil}). The arrows indicate the specific wave functions that the Hamiltonian operates on. By diagonalizing the Hamiltonian (\ref{Hamil}), we obtain wave functions in the three regions of Fig. \ref{fig1} (left and right superconducting regions, and the middle normal regions). The associated wave functions in the superconducting regions for two specific cases are presented in the Supplemental Material \cite{sm}. The analytical expressions for the wave functions in the presence of an arbitrary displacement are rather complicated and we evaluate them numerically. 

Next, we match the wave functions at the superconductor-normal interfaces ($x=\pm d/2$) in Fig. \ref{fig1} and consider a situation where $W\gg d$ so that the lateral edge effects are negligible. Further details of the numerical algorithm employed for studying the supercurrent can be found in Refs. \onlinecite{BP1-prb2018,BP2-prb2018,WS-prb2020,PSH-prb2020}. This method agrees well with Gorkov's Green's function approach and also can provide access to the quasiclassical regime where the Fermi energy is the largest energy in the system \cite{BP2-prb2018,WS-prb2018,prb2016,prb2017}. To simplify the calculations, we have ignored the inverse proximity effects and considered a situation where the roughness of the interfaces plays a relatively weak role, which is the experimentally relevant regime \cite{RMP-2009-Neto,RMP-2008-Beenakker}. Finally, we substitute the numerical wave functions into the current equation above and evaluate the supercurrent as a function of the superconducting phase difference between the two superconducting regions $\varphi=\varphi_l-\varphi_r$.
\begin{figure*}[t]
\includegraphics[ width=17.50cm,height=7.3cm]{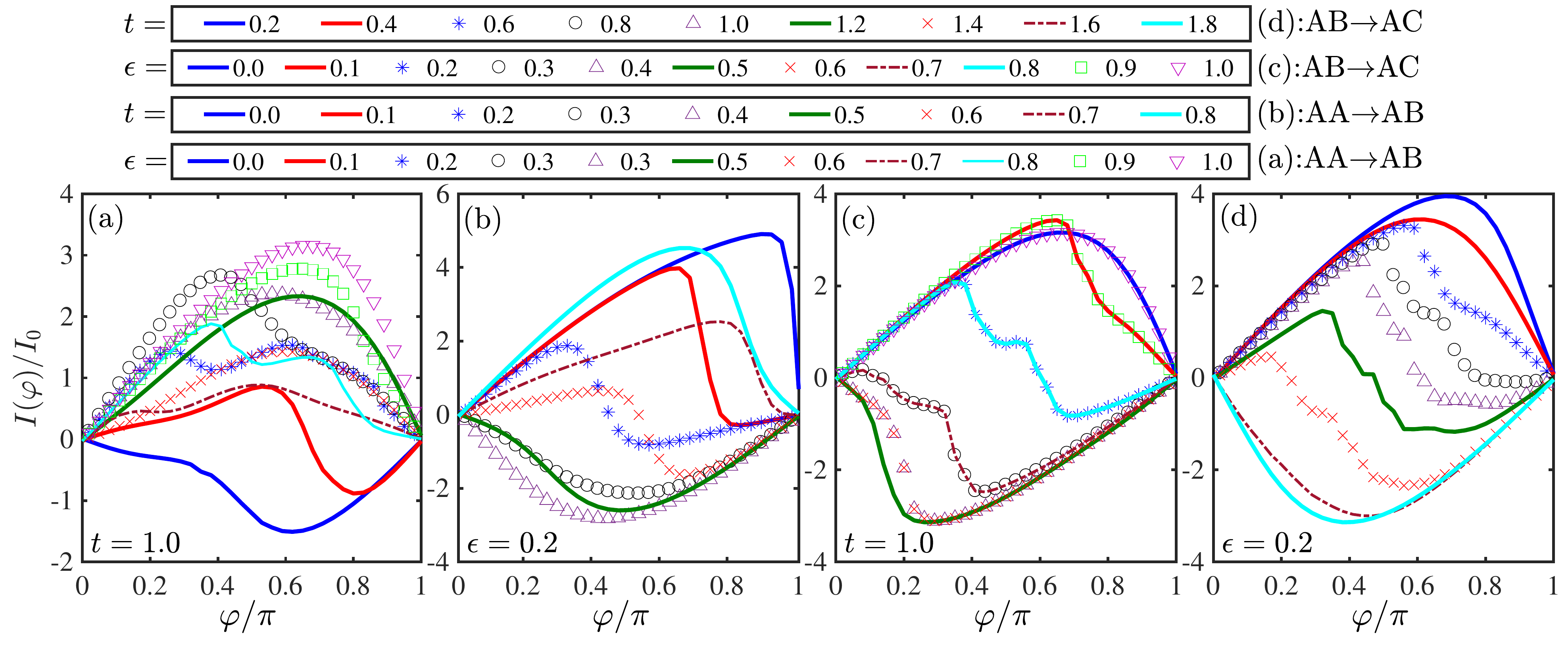}
\caption{\label{fig3} (Color online). The normalized supercurrent as a function of superconducting phase difference for interlayer electron-electron pairing. The initial stacking orders in (a)-(b) and (c)-(d) are AA and AB, respectively. In (a) and (c), the orderings evolve into AB and AC, respectively, by increasing $\epsilon$ from $0$ to $1$. In (b) and (c), the displacement is fixed at $\epsilon=0.2$ and the interlayer coupling parameter varies. }
\end{figure*}

We next consider some experimental aspects. The Josephson junction can be created, for example, by removing a region of thickness $d$ with lithography from the BLG system, or by depositing gate electrodes on BLG to control and pass the supercurrent. As a recent experiment demonstrated \cite{X.Lu}, a so-called `tear-and-stack co-lamination' technique can be employed to produce an accurately defined twisted BLG. Therefore, to test our findings, one feasible fashion is to create several Josephson junction samples with various incrementally displaced layers with the possibility of out-of-plane pressure exertion \cite{M.Yankowitz}. The out-of-plane pressure can be achieved, for example, by compressing a noble gas into a chamber where the sample is placed \cite{M.Yankowitz}. Interestingly, these experiments reported an observation of a Fraunhofer response, characteristic of a Josephson junction, 
the origin of which is currently under investigation \cite{pvcomm}.

Figure~\ref{fig2} shows the supercurrent flow as a function of the superconducting phase difference when superconductivity arises due to an intralayer opposite-spin electron-electron coupling. In Figs.~\ref{fig2}(a) and \ref{fig2}(c), the displacement factor changes from $0$ to unity by a step of $0.1$, transforming AA and AB orderings into AB and AC, respectively (labeled in the legends), and the coupling constant is set at a representative value: $t/\Delta=1$. As seen in Fig.~\ref{fig2}(a), transforming the stacking order from AA to AB causes no supercurrent reversal. However, Fig.~\ref{fig2}(c) shows that evolving from AB to AC order, the supercurrent reverses its direction of flow, and that higher harmonics appear. This effect is prominent even at a relatively small displacement: $\epsilon=0.2$. By tuning the displacement to $\epsilon=0.5$, the reversed supercurrent reaches a maximum. Displacing further the layers towards AC ordering, the supercurrent undergoes another reversal at values greater than $\epsilon=0.8$. To illustrate how the strength of interlayer coupling $t$ influences the supercurrent, we choose $\epsilon=0.2$ and plot the supercurrent for various values of $t$ in Figs.~\ref{fig2}(b) and \ref{fig2}(d). The results show that the AA stacking does not induce a supercurrent reversal in a wide range of interlayer coupling strengths. On the other hand, the supercurrent in AB ordering is sensitive to $t$ so that the weakening or strengthening of interlayer coupling, measured from $t/\Delta=1$, generates a clear supercurrent reversal. The interlayer coupling strength is controllable by introducing tensile or compressive strain perpendicular to the plane of the bilayer. Therefore, in addition to the in-plane displacement, the interlayer coupling can serve as an experimentally adjustable geometric parameter for controlling the direction of supercurrent flow. 
\\
\par
The results of a similar investigation where superconductivity next originates from an interlayer electron-electron coupling are presented in Fig.~\ref{fig3}. All parameter values are the same as in Fig.~\ref{fig2} unless otherwise stated. The supercurrent now changes its direction for both initially AA and AB ordered structures (Fig.~\ref{fig3}(a)-\ref{fig3}(b) and Figs.~\ref{fig3}(c)-\ref{fig3}(d), respectively). The origin of this behavior can be traced back to the actual forms of the Green's function components. As was found in Ref. \cite{alidoust2019prb}, in the AA ordering with intralayer electron-electron coupling, an equal-(pseudo)spin interlayer odd-parity superconducting correlation can appear, which is absent in other cases. Integrating the interlayer odd-parity correlation over the configuration space, the contribution of this component in total supercurrent vanishes. Also, we emphasize that the supercurrent reversals found above disappear for on-site spin-singlet electron-electron couplings. {\color{black} The behavior of the supercurrent is governed by the complicated interplay of lattice symmetry and the actual pairing mechanism. If unconventional odd-parity $p$-wave and $f$-wave pairing mechanisms are involved, the combination of an appropriate Zeeman field in a Josephson junction with different $p$-($f$-)wave orientations in superconducting leads (or $s$-wave electrodes) can result in a self-biased current \cite{BP1-prb2018,BP2-prb2018,WS-prb2020,WS-prb2018,PSH-prb2020}}. 
The cases considered above do not exhaust all theoretical possibilities for introducing superconductivity in the BLG system, but they serve as illustrative examples for what can take place, while further computations should be guided by experimental systems. {\color{black} In terms of realistic values, as seen in Figs. \ref{fig2}(c) and \ref{fig3}(a), by the application of the in-plane displacement of $|\textbf{u}_1|=\epsilon a=0.1\times 1.42 {\AA}=0.142{\AA}$, the supercurrent starts to change direction and goes under a $0$-$\pi$ crossover. The effective model Hamiltonian we consider in this Rapid Communication, can support in-plane and out-of-plane displacement as well as commensurate in-plane twist. Nevertheless, one can expect similar responses for the supercurrent in an incommensurate in-plane twist as one deals with the interplay of lattice symmetry and the pairing mechanism in this scenario, too \cite{AFM}.} 

In summary, we have studied the supercurrent behavior in a Josephson junction configuration based on a BLG system where the pristine graphene layers have changing displacements with respect to each other. We find that the supercurrent in a junction hosting interlayer superconductivity undergoes a reversal upon the application of an out-of-plane strain or in-plane displacement of the layers with AA and AB initial stacking. This effect, however, is absent in the intralayer superconductivity scenario with an initial AA ordering, similar to the on-site spin-singlet superconductivity. We suggest that the behavior of the supercurrent can be exploited to determine the pairing mechanism in BLG systems, which should be a complex interplay of a pairing mechanism with lattice symmetries. Moreover, the richness of the supercurrent behavior with multiple current reversals suggests that BLG Josephson junctions can be considered as attractive candidates for future superconducting memory devices where high-speed switching $0$-$\pi$ crossovers can replace conventional $0$-$1$ computer bits\cite{qbit1}. 

\begin{acknowledgements}
Center for Nanostructured Graphene (CNG) is supported by the Danish National Research Foundation (Project No. DNRF103).
\end{acknowledgements}

\twocolumngrid

\end{document}